\documentclass{vgtc}                          

\graphicspath{{figures/}{pictures/}{images/}{./}} 

\usepackage{times}                     

\usepackage{tabu}                      
\usepackage{booktabs}                  
\usepackage{lipsum}                    
\usepackage{mwe}                       
\usepackage{amsmath}
\usepackage{mathptmx}                  
\usepackage{kotex}
\usepackage{listings}
\usepackage{fancyvrb}

\renewcommand{\paragraph}[1]{\vspace{4pt}\noindent\textbf{#1.}}

\usepackage{booktabs}
\usepackage{multirow}

\usepackage{tikz}
\newcommand{\circled}[1]{%
  \tikz[baseline=(char.base)]{%
    \node[shape=circle, fill=black, text=white, inner sep=1pt, font=\footnotesize\bfseries] (char) {#1};%
  }%
}

\newsavebox{\FVerbatimbox}
\newlength{\FVerbatimwidth}
\AtBeginDocument{%
  \sbox0{\ttfamily X}%
  \setlength{\FVerbatimwidth}{60\wd0}%
  \addtolength{\FVerbatimwidth}{2\fboxsep}
}

\onlineid{1013}

\vgtccategory{Research}

\vgtcinsertpkg

\title{Physical Containers as Framing Conditions for Visualization\\in Augmented Reality}

\author{
Jiyeon Bae$^{1}$\thanks{e-mail: jybae@hcil.snu.ac.kr}
\and Mingyu An$^{1}$\thanks{e-mail: mgahn0706@snu.ac.kr}
\and Jeongin Park$^{1}$\thanks{e-mail: jipark@hcil.snu.ac.kr}
\and Seokweon Jung$^{2}$\thanks{e-mail: skwnjung@kaist.ac.kr}
\and Kiroong Choe$^{1}$
\and Jinwook Seo$^{1}$\thanks{e-mail: jseo@snu.ac.kr}
}

\affiliation{
\vspace{-9pt}
$^{1}$Seoul National University, Republic of Korea\\
$^{2}$KAIST, Republic of Korea
}

\teaser{
    \vspace{-1.5em}
  \centering
  \includegraphics[width=\linewidth]{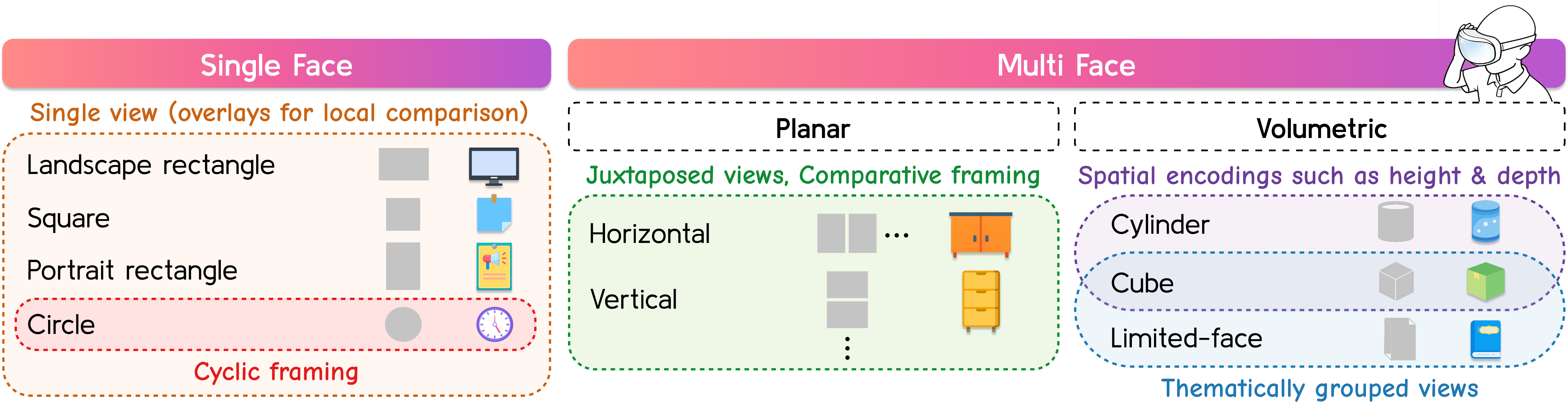}
  \vspace{-2em}
\caption{Physical containers classified by the number of embeddable faces (single vs.\ multi) and form (planar vs.\ volumetric), with associated perceptual framing tendencies for visualization in Augmented Reality (AR).}
  \label{fig:teaser}
}

\abstract{ Exploratory data analysis (EDA) is often hindered by cold-start friction; when users lack specific analytic goals, they struggle to configure complex visualization parameters. While existing visualization tools mostly rely on explicit user input to frame data, we propose leveraging the physical environment as an implicit framing mechanism. We introduce a conceptual framework that uses the geometric and spatial properties of physical containers in Augmented Reality (AR) to guide data interpretation. We characterize how container attributes, such as number of faces, size, proportion, and shape, give rise to distinct perceptual tendencies. For example, a circular container may encourage cyclic interpretation, while juxtaposed planar faces may facilitate comparative analysis. By treating physical forms as environmental framing conditions, we show how AR can orient a user's attention and structure their exploration without requiring manual encoding or prescribing fixed conclusions. We demonstrate this framework through a series of AR design examples illustrating how container morphology foregrounds cyclic, comparative, and sequential analytic patterns.
} 


\begin{document}


\firstsection{Introduction}
\maketitle

Exploratory data analysis (EDA) involves the discovery of patterns and anomalies prior to the formulation of specific analytic questions~\cite{tukey1977}. It often begins before users know what to look for, relying on open-ended visual inspection to form initial impressions. In the absence of explicit guidance, users tend to focus on visually salient features—such as prominent trends or peaks—that may not align with the visualization’s intended analytic goals~\cite{Quadri24}.

We define framing as the perceptual conditions that orient attention and structure interpretation before explicit analytic intent is formed. Under this definition, most visualization systems treat framing as an explicit configuration: users specify axes, encodings, or filters upfront to structure their analysis. While effective when users already have clear analytic goals, this workflow presumes that they know how to look—creating a circular dependency that can stall exploration before interpretive framing begins.

To address this issue, we treat framing not as a user-specified configuration but as an environmental condition. Augmented reality (AR) is particularly well suited to this problem because it allows visualizations to be embedded within physical environments whose geometry can shape interpretation. Prior work suggests that spatial layout and embodied interaction can influence perception and recall~\cite{Niklas23,Hurter24}. Whereas prior studies have examined these effects as consequences of immersive presentation, we elevate them into an explicit design perspective by relating physical container properties—such as number of faces, size, proportion, and shape—to distinct interpretive tendencies. These tendencies influence how many views are encountered concurrently, at what level of granularity data is interpreted, and whether linear or cyclic schemas are activated. This work shifts the unit of analysis in immersive analytics from the visualization itself to the physical structures that frame it.

\section{Background and Related Work}

\subsection{Visualization Interpretation Beyond Guided Tasks}
Prior work has shown that, in the absence of explicit analytic tasks or instructions, viewers’ interpretations of visualizations can diverge from designers’ intended messages, as attention is drawn to salient features shaped by perceptual cues such as size, position, and contrast~\cite{Quadri24}. Cibulski et al.~\cite{cibulski25} further suggest that early encounters with visual information occur before analytic goals or evaluative criteria are fully specified. Together, these findings highlight the importance of framing in early-stage exploration, where interpretation emerges before users know what to look for.

\subsection{Spatial Extensions of Visualization}
Prior work suggests that spatial extensions of visualization are perceptually consequential rather than merely representational. Immersive environments can provide perceptual cues that influence how spatial encodings are perceived~\cite{whitlock2020}, and mixed reality conditions have been shown to enhance users’ recall of content and spatial arrangement compared to traditional 2D and virtual reality settings~\cite{Hurter24}. Patnaik et al.~\cite{Patnaik24} show that users appropriate physical objects and surfaces as semantic proxies for organizing and recalling analytic information, even when the data itself is not inherently situated. However, these studies do not treat the geometric properties of physical containers as framing conditions for visualization. Our work builds on this line of research by characterizing how container properties---face count, size, proportion, and shape---can guide interpretation during early-stage AR exploration.
\section{Containers as Perceptual Framing Conditions}

\subsection{Physical Properties as Given Conditions}
In AR environments where physical objects serve as containers for embedding visualizations, framing can emerge from the geometric properties of those objects rather than from user-specified configuration. Physical containers differ along several fundamental properties, including the number and configuration of faces, size and proportion, and shape. These properties are not specified during analysis, but are already in place when a visualization is encountered, shaping how it can be viewed before analytic intent is formed.

\begin{figure}[tb]
 \centering
 \includegraphics[width=\columnwidth]{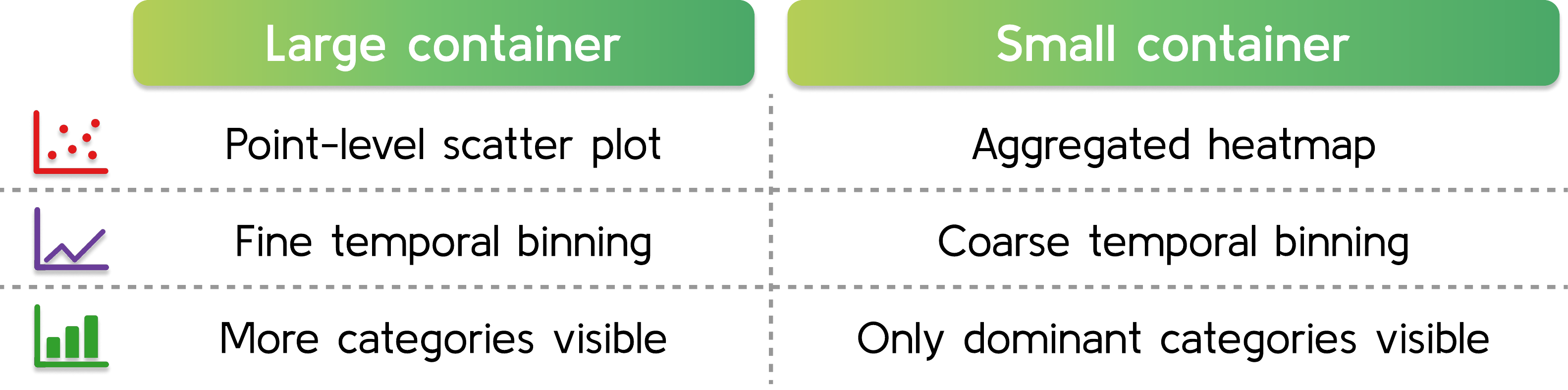}
\vspace{-0.6cm}
 \caption{Container size can influence data granularity across spatial, temporal, and categorical examples: larger containers afford finer detail, while smaller containers encourage aggregation.}
 \vspace{-0.6cm}
 \label{fig:table}
\end{figure}

The number and configuration of faces constrain how many visual surfaces can be perceived simultaneously, influencing whether visual elements are encountered in isolation or in relation to one another. Whether a container can be repositioned or rotated further affects how many of its faces are practically accessible. Size and proportion regulate the spatial extent available for rendering visual marks, implicitly setting limits on granularity, aggregation, and density~(Figure~\ref{fig:table}). Shape further conditions perceptual framing by influencing the directionality and continuity with which visual elements are scanned. Unlike traditional interfaces, where framing results from deliberate layout or encoding choices, container geometry establishes framing prior to analysis.

\subsection{From Conditions to Framing Tendencies}
We focus on three ways in which physical container properties shape perceptual framing: (1) whether a single view or multiple views are perceived concurrently, (2) at what level of granularity data is interpreted, and (3) whether linear or cyclic schemas are activated. While these framing conditions do not prescribe specific analytic goals or conclusions, they shape how data is encountered and interpreted, increasing the salience of certain patterns or relationships. Figure~\ref{fig:teaser} summarizes representative container types and their associated framing tendencies.

These framings should be understood not as fixed design rules, but as tendencies that arise from the interaction between container properties and human perception. The same data can support different framings when embedded in different containers.
\section{Design Examples}

\begin{figure}[tb]
 \centering
 \includegraphics[width=\columnwidth]{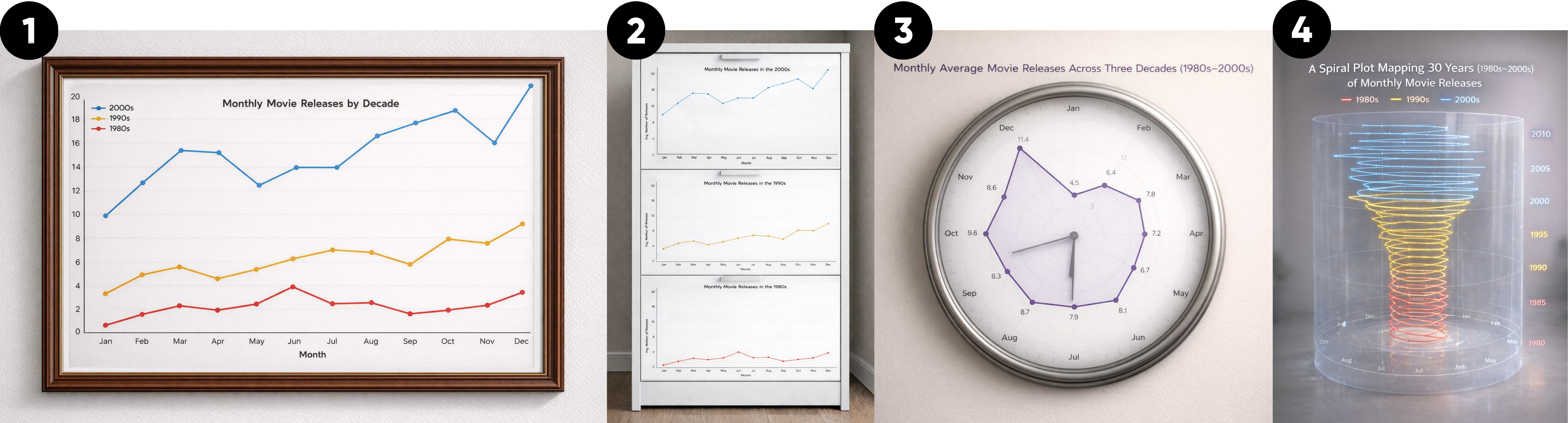}
\vspace{-0.6cm}
\caption{Four AR design examples embedding the monthly movie release counts over three decades into different physical containers. Each activates a different perceptual framing tendency.}
 \vspace{-0.6cm}
 \label{fig:example}
\end{figure}

We use a single dataset—monthly movie release counts over three decades—across examples, allowing framing differences to be attributed to container form rather than the underlying data (Figure~\ref{fig:example}).

\vspace{-0.06cm}

\paragraph{\circled{1} Single-Face · Landscape Rectangle · Overlaid Lines} 
An overlaid line chart embedded on a single rectangular face facilitates comparison across decades, making decade-level growth recognizable at a glance.

\vspace{-0.06cm}

\paragraph{\circled{2} Multi-Face · Planar · Juxtaposed Line Charts}
By separating line charts into juxtaposed views, the design shifts comparison from cross-series overlay to within-series temporal structure, enabling clearer identification of peaks, troughs, and local variation within each decade as an independent series.

\vspace{-0.06cm}

\paragraph{\circled{3} Single-Face · Circle · Polar Line Chart}
The circular container provides twelve angular positions, one per month, which shapes the representation toward aggregation of monthly values across years.
Mapped onto a polar line chart, this geometry foregrounds the cyclic structure, making seasonal patterns and relative peak months more salient than absolute magnitude differences.

\vspace{-0.06cm}

\paragraph{\circled{4} Multi-Face · Volumetric · Helical Time-Series}
Mapping 30 years of monthly data onto a spiral trajectory on a cylindrical surface separates seasonal cycles from long-term accumulation, enabling viewers to perceive recurring monthly patterns alongside gradual growth across years as a sequential temporal flow.

\section{Limitations and Future Work}

The taxonomy in Figure~\ref{fig:teaser} covers container types but is not exhaustive. Containers whose surfaces would require warping a planar visualization may introduce representational distortion and were therefore beyond the scope of this work. Extending the framework to accommodate composite and asymmetric objects and to map them to framing tendencies remains an important direction for future work.

While the present work offers a conceptual account of how container properties shape perceptual framing, controlled user studies are needed both to examine whether the proposed framing tendencies are consistently activated in practice and to isolate the perceptual contribution of container geometry from that of the encoding itself.
Ultimately, grounding these effects empirically may enable the design of AR systems that reduce exploratory friction by leveraging environmental cues rather than explicit configuration.



\bibliographystyle{abbrv-doi}

\bibliography{template}

@inproceedings{Quadri24,
author = {Quadri, Ghulam Jilani and others},
  title = {Do You See What I See? A Qualitative Study Eliciting High-Level Visualization Comprehension},
  year = {2024},
  booktitle = {Proceedings of the 2024 CHI Conference on Human Factors in Computing Systems},
  doi = {10.1145/3613904.3642813}
}

@ARTICLE{Hurter24,
  author={Hurter, Christophe and others},
  journal={IEEE Transactions on Visualization and Computer Graphics}, 
  title={Memory Recall for Data Visualizations in Mixed Reality, Virtual Reality, 3D and 2D}, 
  year={2024},
  volume={30},
  number={10},
  pages={6691-6706},
  doi={10.1109/TVCG.2023.3336588}
}

@article{Niklas23,
  author  = {Elmqvist, Niklas},
  title   = {Data Analytics Anywhere and Everywhere},
  journal = {Communications of the ACM},
  year    = {2023},
  volume  = {66},
  number  = {12},
  pages   = {52--63},
  doi     = {10.1145/3584858}
}

@inproceedings{Patnaik24,
  author    = {Patnaik, Biswaksen and Peng, Huaishu and Elmqvist, Niklas},
  title     = {VisTorch: Interacting with Situated Visualizations using Handheld Projectors},
  booktitle = {Proceedings of the 2024 CHI Conference on Human Factors in Computing Systems},
  year      = {2024},
  doi       = {10.1145/3613904.3642857}
}

@misc{cibulski25,
      title={Towards Understanding Decision Problems As a Goal of Visualization Design}, 
      author={Lena Cibulski and Stefan Bruckner},
      year={2025},
      eprint={2507.18428},
      archivePrefix={arXiv},
      primaryClass={cs.HC},
      url={https://arxiv.org/abs/2507.18428}, 
}

@book{tukey1977,
  title={Exploratory Data Analysis},
  author={Tukey, J.W.},
  year={1977},
  publisher={Addison-Wesley Publishing Company}
}

@inproceedings{Whitlock2020,
  author    = {Whitlock, Matt and Smart, Stephen and Szafir, Danielle Albers},
  title     = {Graphical Perception for Immersive Analytics},
  booktitle = {2020 IEEE Conference on Virtual Reality and 3D User Interfaces (VR)},
  year      = {2020},
  pages     = {616--625},
  doi       = {10.1109/VR46266.2020.00084}
}
\end{document}